\documentclass[jtwocolumn,article,10pt]{IEEEtran}

\ifCLASSINFOpdf
  \usepackage[pdftex]{graphicx}
\else
  \usepackage[dvips]{graphicx}
\fi

\usepackage[cmex10]{amsmath}
\usepackage{amssymb}
\usepackage{float}

\usepackage{bm}
\usepackage{extarrows}

\usepackage{algorithm}
\usepackage{algorithmic}

\usepackage{stfloats}

\makeatletter

\newcommand{\Rmnum}[1]{\expandafter\@slowromancap\romannumeral #1@}
\makeatother

\usepackage{footnote}
\makesavenoteenv{tabular}

\usepackage{graphicx,color}
\begin{document}
\vspace{-0.5cm}

\title{Study of BEM-Type Channel Estimation Techniques for 5G Multicarrier Systems}

\author{\IEEEauthorblockN{Joao T. Dias$^{1,2}$, 
Rodrigo C. de Lamare$^{2,3}$ and 
 Yuriy V. Zakharov$^{3}$} \\
\vspace{5mm}
\IEEEauthorblockA{$^{1}$Federal Center for Technological Education of Rio de Janeiro (CEFET-RJ)},\\
$^{2}$Centre for Telecommunications Studies (CETUC) of Pontifical Catholic University of Rio de Janeiro, Brazil and\\
$^{3}$Department of Electronic Engineering, University of York, UK\\
e-mail: joao.dias@cefet-rj.br, $\left\{ \text{joao.dias, delamare}\right\}$@cetuc.puc-rio.br and $\left\{ \text{rodrigo.delamare, yury.zakharov}\right\}$@york.ac.uk}
\maketitle

\thispagestyle{empty}

\begin{abstract} ~ In this paper, we investigate channel estimation techniques for 5G multicarrier systems. Due to the characteristics of the 5G application scenarios, channel estimation techniques have been tested in Orthogonal Frequency Division Multiplexing (OFDM) and Generalized Frequency Division Multiplexing (GFDM) systems. The orthogonality between subcarriers in OFDM systems permits inserting and extracting pilots without interference. However, due to pulse shaping, subcarriers in GFDM are no longer orthogonal and interfere with each other. Due to such interference, the channel estimation for GFDM is not trivial.  A robust and low-complexity channel estimator can be obtained by combining a minimum mean-square error (MMSE) regularization and the basis expansion model (BEM) approach. In this work, we develop a BEM-type channel estimator along with a strategy to obtain the covariance matrix of the BEM coefficients. Simulations show that the BEM-type channel estimation shows performance close to that of the linear MMSE (LMMSE), even though there is no need to know the channel power delay profile, and its complexity is low.
\end{abstract}


\IEEEpeerreviewmaketitle

\section{Introduction}

The fifth generation (5G) of mobile systems must handle diverse
scenarios and meet the increasing  demand for high data rates. To
achieve an efficient use of limited spectrum, Cognitive Radio (CR)
has gained a great deal of interest.  The idea is to allow the
exploitation of locally available frequency bands, on a temporary
basis but under a non-interfering constraint. In order to avoid the
interference from an opportunistic user to the primary user, a
waveform used by the secondary user must have a low out of band
radiation (OOB)~\cite{Nicola}. Most modern digital communication
standards use Orthogonal Frequency Division Multiplexing (OFDM)
\cite{Ahmad} as the air interface, because of its flexibility and
robustness in frequency-selective channels. Nevertheless, OFDM
presents some drawbacks such as the high OOB \cite{Jaap} that
affects its application especially in CR systems.
In contrast, GFDM can achieve a low OOB using a spectrally-contained pulse on each subcarrier, and it has higher spectrum efficiency, because it does not need to use virtual subcarriers to avoid adjacent channel interference, and because it reduces the ratio between the guard time interval and the total frame duration~\cite{Nicola, Dias}.
However, the subcarrier filtering results in non-orthogonal subcarriers, leading to inter-symbol interference (ISI) and inter-carrier interference (ICI).

In this work, we develop a basis expansion model (BEM) approach for
estimation of the channel parameters~\cite{Edfor, Zakharov, Yuri} in
OFDM and GFDM systems. We first consider classical least-squares
(LS) and linear minimum mean-square error (LMMSE), and then examine
BEM-type channel estimation algorithms. We then use a strategy to
estimate the covariance matrix of the channel vector containing the
coefficients of the channel and incorporate it into a BEM-type
channel estimation technique that approximates the LMMSE channel
estimator. Simulations show that the BEM estimator presents results
close to those of the LMMSE estimator, but there is no need to know
the power delay profile (PDP) of the channel and the estimator
complexity is feasible.

The paper is organized as follows: in Section II, a description of
the signal model of the OFDM and GFDM systems is given; Section III
describes estimation algorithms known in the literature; the BEM
algorithms are described in Section IV; Section V describes the
covariance matrix approach and aproximated-LMMSE-BEM algorithm; the
results of the simulations are presented in Section VI and in
Section VII, some conclusions are drawn.

\section{Signal Model}
\label{ProbState}

The OFDM transmit signal is given by \cite{Ahmad}
\begin{equation} \label{eq:ofdmtransmit}
x_{\text{OFDM}} [n]=\displaystyle \sum_{k=0}^{K-1} d_{k}e^{j2\pi \frac{k}{K}n},
\end{equation}
where $d_{k}$ is the data symbol at the $k^{th}$ subcarrier and $K$
is the number of subcarriers in the OFDM symbol.



In GFDM, each data symbol is pulse-shaped using a circularly time
and frequency shifted version of a prototype filter $g[n]$ whose
energy over one period $N=MK$ is normalized to one ($M$ is the
number of subsymbols per GFDM block and $K$ is the number of
subcarriers per subsymbol). The GFDM transmit signal is given by
\cite{Nicola}
\begin{equation} \label{eq:gfdmtransmit}
x_{\text{GFDM}} [n]=\displaystyle \sum_{k=0}^{K-1}\displaystyle \sum_{m=0}^{M-1}d_{k}[m] g[(n-mK)~ \text{mod}~ N]e^{j2\pi \frac{k}{K}n},
\end{equation}
where $d_{k}[m]$ is the data symbol at time index $m$ and subcarrier index $k$.

\begin{figure*}[!th]
\centering
\includegraphics[width=.8\textwidth]{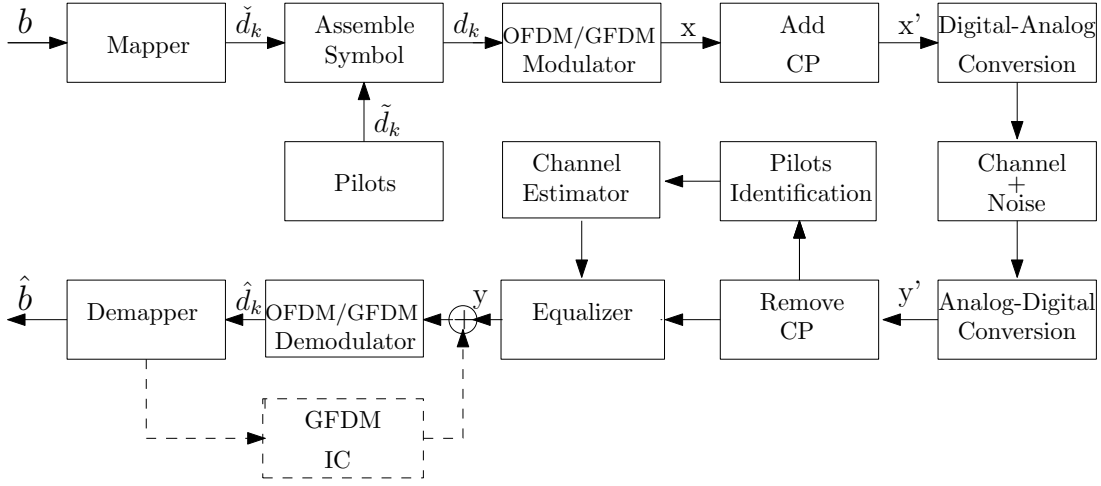}
\caption{Block diagram of the proposed solution for the OFDM and the GFDM systems}
\label{block_diagram_ofdm_gfdm}
\end{figure*}
In order to insert and extract interference-free pilots, we rewrite the GFDM transmit signal by \cite{Shahab}
\begin{equation} \label{eq:gfdmtransmit}
\bold{x}_{\text{GFDM}}=\bold{W}_N^H\displaystyle \sum_{k=0}^{K-1}\bold{P}^{(k)}\bold{G}^{(\delta)}\bold{S}^{(\delta)} \left(\bold{\Gamma} \tilde{\bold{d}}_k+\bold{W}_M \check{\bold{d}}_k \right)  \in \mathbb{C}^{N  \times 1},
\end{equation}
where $\check{\bold{d}}_k$ contains the symbols at data subcarriers with zeros at pilot subcarriers and $\tilde{\bold{d}}_k$ carries information at only the pilot subcarriers with zeros at data subcarriers.  
$\bold{W}_M$ is the unitary DFT matrix of size $M\times M$ and $\bold{\Gamma}=\bold{P'}\text{blkdiag}(\lambda\bold{I}_{n_p},\bold{W}_{M-n_p})$. Here, $\bold{P'}$ can be any permutation matrix of compatible size which allocates the pilots to any frequency bin within the pilot subcarriers, and $\text{blkdiag}(\bold{X},...,\bold{Y})$ is a block diagonal matrix according to its matrix entries with $\bold{X}$ being the top-left and $\bold{Y}$ being the bottom-right blocks. The parameter $\lambda$ is a scaling factor that normalizes the pilot energy to one, and $\bold{I}_{n_p}$ is an identity matrix of size $n_p$, where $n_p$ is the number of pilots in each $\tilde{\bold{d}}_k$. $\bold{S}^{(\delta)}=\bold{1}_{\delta,1}\otimes\bold{I}_M$ is $\delta$-fold repetition matrix which concatenates $\delta$ identity matrices of size $M$; $\bold{1}_{i,j}$ is a $i\times j$ matrix of ones, and $\otimes$ is the Kronecker product. The value of $\delta$ is based on the number of non-zero values in the filter frequency response. The subcarrier filter $\bold{G}^{(\delta)}=\text{diag}(\bold{W}_{M\delta}\bold{g}^{\delta})$ is diagonal in the frequency domain. The circulant filter $\bold{g}^{\delta}$ is the version of $\bold{g}=(g[n])_{n=0,...N-1}$ down-sampled by factor $K/\delta$. An up-conversion of the $k^{th}$ subcarrier to its respective subcarrier frequency is performed by the shift matrix $\bold{P}^{(k)}=\bold{\Psi}(\bold{p}^{(k)})\otimes\bold{I}_M$, where $\bold{\Psi}(.)$ returns the circulant matrix based on the input vector, and $\bold{p}^{(k)}$ is a column vector where the $k^{th}$ element is $1$ and all others are zeros. The K subcarriers are summed and transformed back to the time-domain with $\bold{W}_N^H$, where $(.)^H$ is the Hermitian transpose of (.).

To ensure the interference-free pilot insertion, in expression
(\ref{eq:gfdmtransmit}) the plus sign cannot superimpose the
information, i.e., if $k$ is a pilot subcarrier $\bold{W}_M
\check{\bold{d}}_k$ becomes $\bold{0}_M$ and if it belongs to data
subcarriers $\bold{\Gamma} \tilde{\bold{d}}_k$, it would be
$\bold{0}_M$. The pilots need to be located at the frequency bins
where no inter-carrier interference is present.

At the receiver, the received signal is described by
\begin{equation} \label{eq:signalreceived}
\bold{y}'=\bold{H}\bold{x}' +\bold{n}  \in \mathbb{C}^{(KM+L)\times 1}
\end{equation}
where $\bold{H} \in \mathbb{C}^{(KM+L)\times (KM+L)}$ describes the channel, $\bold{x}'$ is the vector $\bold{x}$ with addition of the cyclic prefix (CP), and $\bold{n}$ is a noise vector containing additive white Gaussian noise (AWGN) samples.

The signal is converted from the analog to the digital domain, followed by removal of the CP. Then, we apply 
the Fourier transform to extract the pilots in the frequency domain,
to estimate the channel $\bold{\hat{H}}$ and to obtain
$\bold{y}=\bold{\hat{H}}^{-1}\bold{y}'$.

On the demodulator side, the recovered data symbols for the $k^{th}$ subcarrier in GFDM system are given by
\begin{equation} \label{eq:gfdmreceive}
\bold{\hat{d}}_{k}=\bold{W}_M^H(\bold{S}^{(\delta)})^T(\bold{G}^{(\delta)})^{\ast}(\bold{P}^{(k)})^T \bold{W}_N \bold{y} ~~ \in \mathbb{C}^{M  \times 1}.
\end{equation}

In order to mitigate the interference between adjacent data
subcarriers, we use the interference cancellation (IC) algorithm
proposed  in \cite{Datta}. Note that other interference cancellation
techniques
\cite{delamare_mber,rontogiannis,delamare_itic,stspadf,choi,stbcccm,FL11,delamarespl07,jidf,jio_mimo,tds,peng_twc,spa,spa2,jio_mimo,P.Li,jingjing,memd,did,bfidd,mbdf,bfidd,mserrr,mmimo,wence,shaowcl08}.
can also be examined. In this case, the notation needs to be
extended by an iteration index $j=1,...,J$. The vector of received
frequency domain samples $\bold{y}_k$ then corresponds to
$\bold{y}_k^{(j)}$, and the received data symbols
$\bold{\hat{d}}_{k}$ to $\bold{\hat{d}}_{k}^{(j)}$, respectively.
With this notation, $\bold{y}_k^{(0)}$ and
$\bold{\hat{d}}_{k}^{(0)}$ denote vectors on which no interference
cancellation has been performed. The IC algorithm is detailed in
Algorithm 1.

\begin{algorithm}[]
\caption{IC}
\begin{algorithmic}[1]
\STATE{Consider $\bold{y}_k=(\bold{S}^{(\delta)})^T(\bold{G}^{(\delta)})^{\ast}(\bold{P}^{(k)})^T \bold{W}_N \bold{y}$}
\STATE{receive all subcarriers as $\bold{W}_{M}^H\bold{y}_k^{(0)}$}
\STATE{map each symbol to the closest Quadrature Phase Shift Keying (QPSK) point to obtain $\bold{\hat{d}}_{k}^{(0)}$}
    \FOR{$j=1:J$}
      \FOR{$k=0:K-1$}
           \STATE{Remove interference by computing: $\bold{y}_k^{(j)}=\bold{y}_k^{(0)}-\bold{P}^{(k)}\bold{G}^{(\delta)}\bold{S}^{(\delta)}\bold{W}_M\left(\bold{\hat{d}}_{(k-1)~ \text{mod}~ K}^{(j-1)}+\bold{\hat{d}}_{(k+1)~ \text{mod}~ K}^{(j-1)}\right)$}
           \STATE{update the received symbol to the closest QPSK point to obtain $\bold{\hat{d}}_{k}^{(j)}$}
     \ENDFOR
    \ENDFOR
\end{algorithmic}
\end{algorithm}

Essentially, the process consists of detecting the data symbols $\bold{\hat{d}}_{k}^{(j)}$ and using them in the $(j+1)$th iteration to compute the interfering signal $\bold{P}^{(k)}\bold{G}^{(\delta)}\bold{S}^{(\delta)}\bold{W}_M\left(\bold{\hat{d}}_{(k-1)~ \text{mod}~ K}^{(j-1)}+\bold{\hat{d}}_{(k+1)~ \text{mod}~ K}^{(j-1)}\right)$, which is then subtracted from the original signal $\bold{y}_k^{(0)}$ in order to obtain an interference-reduced version of the data estimate from $\bold{y}_k^{(j)}$.

The analysis of the possible error propagation issue in the interference cancellation performed by algorithm 1 is beyond the scope of this work. However, it was possible to observe in the simulations carried out that such error propagation can be considered negligible in the scenario evaluated.

The computational complexity of the systems (OFDM and GFDM) in terms of complex valued multiplications can be expressed as: $\mathcal{O}_{OFDM} (2MK\text{log}_2K)$, where $2$ is due to the fact that this operation occurs in the transmitter  and receiver, $M$ is the number the symbol per block and $K\text{log}_2K$ originates from the $K\times K$ points DFT (see Eq. (\ref{eq:ofdmtransmit})); and $\mathcal{O}_{GFDM} (2(N\text{log}_2N+KLM+KM\text{log}_2M))$, where $KLM$ denotes the matched filtering of the sub-carriers, $N\text{log}_2N$ and $M\text{log}_2M$ originates from the $N\times N$ and  $M\times M$ points DFT, respectively (see Eqs. (\ref{eq:gfdmtransmit}) and (\ref{eq:gfdmreceive})).

Applying the IC algorithm $J$ times to all sub-carriers introduces additional $JKM\text{log}_2M+JKM+JKM\text{log}_2M$ operations for GFDM system, where $JKM\text{log}_2M$ for transforming the estimated data symbols to the frequency domain, $JKM$ for applying the interference filter and another $JKM\text{log}_2M$ for transforming back to the time domain.

The block diagram of the proposed solution for the OFDM and the GFDM systems is shown in Fig.~\ref{block_diagram_ofdm_gfdm}.


 \section{Classical Channel Estimation Algorithms}
 \label{UwProject}
In this section, we describe the channel estimation methods that are utilized for comparison with the BEM based technique. 


 At the receiver side, after performing the time-frequency conversion, we can rewrite~(\ref{eq:signalreceived}) in the pilot positions
 , as:
\begin{equation} \label{eq:signaluwreceived}
\bold{\tilde{y}}_{p}= \bold{\tilde{H}}_{p}\bold{\tilde{x}}_{p} + \bold{\tilde{n}}_{p},
\end{equation}
where $\bold{\tilde{y}}_{p} \in \mathbb{C}^{(N_p)\times 1}$, $\bold{\tilde{H}}_{p} \in \mathbb{C}^{(N_p)\times (N_p)}$, $\bold{\tilde{x}}_p \in \mathbb{C}^{(N_p )\times 1}$ and $N_p$ is the number of pilots per symbol.

\subsection{Least Squares}
Least Squares (LS) \cite{Ozdemir} 
provides the estimate of the channel frequency response at the pilot positions by
\begin{equation} \label{eq:ls}
\bold{\hat{\tilde{h}}_p}_{\text{LS}}= \text{diag}\left\{\bold{\tilde{x}}_{p}\right\}^{-1}\bold{\tilde{y}}_{p},
\end{equation}
where the matrix $(\text{diag}\left\{\bold{\tilde{x}}_{p}\right\}^{-1}) \in \mathbb{C}^{(N_p)\times (N_p)}$ and the vector $\bold{\tilde{x}}_{p}$ occupies the main diagonal of the matrix.

The LS algorithm has low computational complexity $\mathcal{O} (N_p)$,  but the estimates have a high variance due to the noise presence.

\subsection{Linear Minimum Mean Square Error}
 The Linear Minimum Mean Square Error (LMMSE) \cite{Edfor} 
can be written as
 \begin{equation} \label{eq:lmmse2}
 \bold{\hat{\tilde{h}}_p}_{\text{LMMSE}}= \bold{R}_{\tilde{\bold{h}}_{p}\tilde{\bold{h}}_{p}} \left(\bold{R}_{\tilde{\bold{h}}_{p}\tilde{\bold{h}}_{p}} + \frac{\beta}{SNR}\bold{I}_{N_p}\right)^{-1}\bold{\hat{\tilde{h}}_p}_{\text{LS}},
\end{equation}
where $\beta=E\{ |d_k|^2\}E\{\frac{1}{|d_k|^2}\}$, with $d_k$ being the constellation points of the mapping, $SNR$ is the signal to noise ratio, and $\bold{I}_{N_p}$ is an identity matrix.
To compute an element of the channel covariance matrix $\bold{R}_{\tilde{\bold{h}}_{p}\tilde{\bold{h}}_{p}}=E\{\tilde{\bold{h}}_{p}\tilde{\bold{h}}_{p}^H\}$, we need to know the PDP of the channel. 
The need to estimate the channel covariance together with the need to estimate the SNR periodically, makes the use of the LMMSE estimator computationally prohibitive; its computational complexity is $\mathcal{O} (N_p^3)$.

\section{BEM Channel Estimation Algorithms}

The BEM approach represents the channel frequency response using basis functions
\cite{Hrycak}. It should be noted that other subspace techniques
\cite{scharf,bar-ness,pados99,reed98,hua,goldstein,santos,qian,delamarespl07,xutsa,delamaretsp,kwak,xu&liu,delamareccm,wcccm,delamareelb,jidf,delamarecl,delamaresp,delamaretvt,jioel,delamarespl07,delamare_ccmmswf,jidf_echo,delamaretvt10,delamaretvt2011ST,delamare10,fa10,lei09,ccmavf,lei10,jio_ccm,ccmavf,stap_jio,zhaocheng,zhaocheng2,arh_eusipco,arh_taes,dfjio,rdrab,dcg_conf,dcg,dce,drr_conf,dta_conf1,dta_conf2,dta_ls,song,wljio,barc,jiomber,saalt}
exploit the eigenstructure of the input data matrix can be also
considere. Using BEM, the vector $\bold{\tilde{h}}$ of the channel
frequency response can be represented as
\begin{equation}
\bold{\tilde{h}} = \bold{B}\bold{a},
\end{equation}
where $\bold{B}$ is the matrix of basis functions, every column of $\bold{B}$ is a basis function, and $\bold{a}$ is the vector of BEM expansion coefficients.

When the channel estimation is transformed into estimation of the vector $\bold{a}$, its complexity is reduced because the size of $\bold{a}$ is smaller  than $\bold{\tilde{h}}$. 

\subsection{LS-BEM}

To get an estimate $\bold{\hat{a}}$, we can rewrite (\ref{eq:signaluwreceived}) by replacing $\bold{\tilde{h}}$ by  $\bold{B}\bold{a}$ and  we compute $\bold{a}$ that minimizes the cost function $J(\bold{a})= (\bold{\tilde{y}}_{p}- \text{diag}(\bold{\tilde{x}}_{p})\bold{B}\bold{a})^H(\bold{\tilde{y}}_{p}- \text{diag}(\bold{\tilde{x}}_{p})\bold{B}\bold{a})$. This procedure is known as Least-Squares BEM (LS-BEM) \cite{Barhumi}. The minimum of $J(\bold{a})$ is found by differentiating it with respect to $\bold{a}$ and setting the derivative equal to a null vector. The resultant LS estimate of $\bold{a}$ is
\begin{equation}
\bold{\hat{a}} = (\bold{B}^H\bold{\tilde{X}}^H\bold{\tilde{X}}\bold{B})^{-1}\bold{B}^H\bold{\tilde{X}}^H\bold{\tilde{y}}_{p},
\end{equation}
where $\bold{\tilde{X}}=\text{diag}(\bold{\tilde{x}}_{p})$.

Since each pilot subcarrier has modulus equal to $1$ ($|\tilde{\text{x}}_{p}(k)| = 1$), then $\bold{\tilde{X}}^H\bold{\tilde{X}}= \bold{I}$ is an identity matrix. As the matrix $\bold{B}$ with the bases is known a priori, the matrix $\bold{B}^H\bold{B}$ can be precomputed. Its inverse $(\bold{B}^H\bold{B})^{-1}$ can also be precomputed, thus avoiding the real-time matrix inversion.

Taking into account that $\bold{\hat{\tilde{h}}_p}_{\text{LS}}= \text{diag}\left\{\bold{\tilde{x}}_{p}\right\}^{-1}\bold{\tilde{y}}_{p}$, the LS-BEM channel estimate can be represented as
\begin{equation}
\bold{\hat{\tilde{h}}_p}_{\text{LS-BEM}} = (\bold{B}^H\bold{B})^{-1}\bold{B}^H\bold{\hat{\tilde{h}}_p}_{\text{LS}},
\end{equation}
where $(\bold{B}^H\bold{B})^{-1}\bold{B}^H$ can also be precomputed to reduce the real-time computation.

The LS-BEM algorithm has computational complexity $\mathcal{O} (N_p N_a)$, where $N_a$ is the length of the vector $\bold{a}$ which is smaller than $N_p$. However, the LS-BEM channel estimates are sensitive to the additive noise.

\subsection{LMMSE-BEM}

To improve the performance, we take the level of the noise (i.e. the noise variance) into account \cite{Tang}. This is usually done by applying the Bayesian Gauss-Markov theorem~\cite{Kay} to minimize a cost function $J(\bold{\epsilon})= E\{|\bold{\epsilon}|^2\}$, with $\epsilon=\bold{a}-\hat{\bold{a}}$, and by obtaining the LMMSE estimator for $\bold{a}$ as given by
\begin{equation} \label{eq:lmmse3}
 \bold{\hat{a}}=  \left(\bold{B}^H\bold{B} + \sigma_{\tilde{n}}^2\bold{R}_{a}^{-1}\right)^{-1}\bold{B}^H\bold{\tilde{X}}^H\bold{\tilde{y}}_{p},
\end{equation}
where $\bold{R}_{a}= E\{\bold{a}\bold{a}^H\}$ is the covariance matrix of the expansion coefficients.

The covariance matrix of the expansion coefficients can be obtained from the channel covariance, assuming that $\tilde{\bold{h}}_{p} = \bold{B}\bold{a}$, i.e., the modeling error is negligible.
We can find $\bold{R}_{\tilde{\bold{h}}_{p}\tilde{\bold{h}}_{p}} = E\{\tilde{\bold{h}}_{p}\tilde{\bold{h}}_{p}^H\}$. Then, we write $\bold{R}_{\tilde{\bold{h}}_{p}\tilde{\bold{h}}_{p}} = E\{\bold{B}\bold{a}\bold{a}^H\bold{B}^H\} = \bold{B}E\{\bold{a}\bold{a}^H\}\bold{B}^H = \bold{B}\bold{R}_{a}\bold{B}^H$.

By multiplying from the left by $(\bold{B}^H\bold{B})^{-1}\bold{B}^H$ and from the right by $\bold{B}(\bold{B}^H\bold{B})^{-1}$, and also taking into account that $(\bold{B}^H\bold{B})^{-1}\bold{B}^H\bold{B} = \bold{I}$, we obtain
\begin{equation} \label{eq:varch}
\bold{R}_{a} =(\bold{B}^H\bold{B})^{-1}\bold{B}^H\bold{R}_{\tilde{\bold{h}}_{p}\tilde{\bold{h}}_{p}}\bold{B}(\bold{B}^H\bold{B})^{-1}.
\end{equation}

The LMMSE-BEM algorithm has computational complexity $\mathcal{O} (N_p N_a^2)$, but the problem to find the channel PDP is still present.

\subsection{BEM designs}

Many traditional BEM designs have been used to model the channels. In this work, we use the Complex Exponentials (CE-BEM)\cite{Tang} and Legendre Polynomials (LP-BEM)\cite{Hrycak}.

The CE-BEM is described by \cite{Tang}
\begin{equation}  \label{eq:b1}
\bold{B}_{\text{CE}}(q,t)= e^{\frac{-j2\pi(p_s.q) (t-1)}{K}},
\end{equation}
where $\left(\frac{p_s.q}{K}\right)$'s are the normalized frequencies in the pilot positions,  $p_s$ is the pilot separation and $t$'s are the basis function index.

The LP-BEM is described by \cite{Hrycak}
\begin{equation}  \label{eq:b2}
\bold{B}_{\text{LP}}(q,t)= \frac{1}{2^t t!} \frac{d^t}{dq^t}[(q^2-1)^t],
\end{equation}
where $\left(\frac{d^t}{dq^t}\right)$'s are the $t$'s derivatives 
in $q$.

\section{
Covariance Matrix Approach and Aproximated-LMMSE-BEM Algorithm}
\label{ApCovMat}
As known in the literature, the LMMSE regularization requires the knowledge of the channel covariance matrix, which in turn is defined by the channel PDP. It is difficult to accurately estimate it in practice. In order to overcome this problem, we develop a BEM-type channel estimator for GFDM and other multicarrier systems that considers 
that the channel PDP is constant $\frac{1}{L}$ and that the coefficients are evenly distributed along the cyclic prefix (CP), whose length is $L$. 
Therefore, we obtain the covariance matrix for the channel approximation as
\begin{equation} \label{eq:acovmat}
\tilde{\bold{R}}_{\bold{h}'\bold{h}'} (n,p)= \sum \limits_{l=0}^{L-1} \left(\frac{1}{L}\right)e^{\frac{-j2\pi p_s(n-p)(l)}{K}},
\end{equation}
where $n,p = 1, 2,\hdots, N_p$ indicates the position of the pilot subcarrier in the symbol of length $K$. This matrix computation is done once and offline.

The Approximated LMMSE (aLMMSE) is described analogously to (\ref{eq:lmmse3}), through replacing $\bold{R}_{\tilde{\bold{h}}_{p}\tilde{\bold{h}}_{p}}$ by $\tilde{\bold{R}}_{\bold{h}'\bold{h}'} $ in (\ref{eq:varch}).


The aLMMSE-BEM algorithm is detailed in Algorithm 2.

\begin{algorithm}[]
\caption{aLMMSE-BEM}
\begin{algorithmic}[2]
\STATE{$q=1:N_p$}
\STATE{$t=1:$ number~of~basis~functions}
\STATE{Generate $\bold{B}$ with (\ref{eq:b1}) or (\ref{eq:b2})}
    \FOR{$n=1:N_p$}
      \FOR{$p=1:N_p$}
        \FOR{$l=1:L$}
           \STATE{Generate $\tilde{\bold{R}}_{\bold{h}'\bold{h}'} (n,p)$ with (\ref{eq:acovmat})}
       \ENDFOR
     \ENDFOR
    \ENDFOR
\STATE{Calculate (\ref{eq:varch}) by replacing $\bold{R}_{\tilde{\bold{h}}_{p}\tilde{\bold{h}}_{p}}$ by $\tilde{\bold{R}}_{\bold{h}'\bold{h}'} $}
\STATE{Calculate (\ref{eq:lmmse3}) with the new (\ref{eq:varch})}
\STATE{Calculate $\bold{\hat{\tilde{h}}} = \bold{B}\bold{\hat{a}}$ with the vector $\bold{\hat{a}}$ obtained from (\ref{eq:lmmse3})}
\end{algorithmic}
\end{algorithm}

The computational complexity of the aLMMSE-BEM algorithm is $\mathcal{O} (N_p N_a^2)$.

\section{simulations results}
\label{SimResult}
To validate the BEM based channel estimation technique in OFDM and GFDM systems, simulations were performed comparing its performance with that of the LS and LMMSE estimators. The complexity of the solution for each system is the sum of the computational complexity of the system (OFDM or GFDM) added to the computational complexity of the selected channel estimation algorithm. The simulations were performed with the parameters listed in Table~\ref{parameters}.

\begin{table}[h]
\centering
\caption{Simulation Parameters}
\label{parameters}
\vspace{-.5 em}
\begin{tabular}{|l|c|}
  \hline \hline
  number of subcarriers [K] & 128\\ \hline
 pilot separation [$p_s$] & 4 \\ \hline
  GFDM sub-symbols [M] & 5 \\ \hline
  pulse shaping filter [G] & RRC \\ \hline
  roll-off factor [$\alpha$] & $\{0.1, 0.2, 0.5\}$ \\ \hline
  modulation subcarriers& QPSK \\ \hline
  number of CE basis functions [$N_a$]& 18 \\ \hline
  number of LP basis functions [$N_a$]& 18 \\ \hline
  number of iterations of the IC algorithm [$J$]&2 \\ \hline
  \hline
  \end{tabular}
  \end{table}

 The choice of the number of basis functions was performed in order to optimize the MSE in all range of  the tested $E_{b}/N_{0}$.

 In the first example, we observe the  mean square error (MSE) performance in a multipath channel\cite{Zheng}. This channel has been modeled as a tapped-delay line, whose power delay profile is ($1, 0.5, 0.25, 0.125$) for the fractional delays ($0, 2.7, 3.1, 4.9$) samples, each tap is modeled by a Gaussian distribution. 
In order to provide a simple way of equalization, the length of CP was chosen with 8 samples.

\begin{figure}[!h]
\centering
\includegraphics[width=1\columnwidth]{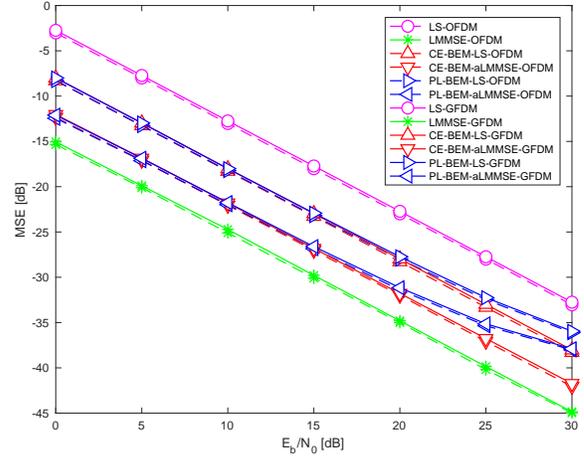}
\vspace{-1em}
\caption{\itshape{MSE performance of channel estimation}}
\label{mse_ofdm}
\end{figure}

In Fig.~\ref{mse_ofdm}, we can observe that the OFDM system (solid line) and the GFDM system (dotted line) show very similar performances, the small difference observed can be attributed to the presence of CP in all OFDM symbols, which degrades the $E_{b}/N_{0}$. The approximated LMMSE BEM  (aLMMSE-BEM) shows a MSE $5$ dB lower than that of the LS-BEM, and  $10$ dB lower than that of the classic LS. The curves CE-BEM-aLMMSE and LP-BEM-aLMMSE show a similar performance for low SNR, but the CE-BEM-aLMMSE shows the best performance for high SNR (above $20$ dB). This  behavior shows that the CE basis functions are a superior choice for the estimation of this channel model, while the LP would need a greater number of basis functions to avoid the floor observed in its MSE estimates. As already known in the literature, the best estimation is obtained with the LMMSE  \cite{Ozdemir}, which requires knowledge of channel PDP and whose complexity is high.

In the second example, shown in Fig.~\ref{ber_ofdm}, we can observe that the Bit Error Rate  (BER) performance for the aLMMSE-BEM estimator is approximately $0.5$ dB better than that of LS-BEM, $1.5$ dB better than that of LS, and very close to that of the LMMSE and when the channel state information (CSI) is known. The CE and LP basis functions show quite similar performance. The best performance of the BER in the GFDM system, when compared to the OFDM system, may be associated with the degradation of the $E_{b}/N_{0}$ due to the presence of the CP in all the OFDM symbols.

\section{Conclusions}
\label{Conclus}
This paper has investigated channel estimation techniques for 5G multicarrier systems. The classical  LS and LMMSE criterion-based estimators have been briefly reviewed and a BEM-type channel estimator has been developed, which effectively approximates the linear MMSE channel estimator. The resulting MSE performance shows that  the aLMMSE-BEM algorithm presents results close to LMMSE, but its computational complexity is reduced. The MSE performance of the GFDM channel estimation shows that the performances are quite similar to OFDM, which shows that the pilots remained orthogonal to the data symbols in the frequency domain, suffering no interference in GFDM systems. The BER curves show that the performance of the proposed estimator is very close to the LMMSE, which requires knowledge of the channel PDP and whose complexity is high.


\section{Acknowledgement}
The work of Y. Zakharov is partly supported by the UK Engineering and Physical Sciences Research Council (EPSRC) through Grant EP/R003297/1.

\bibliographystyle{IEEEtran}
\bibliography{refsrrpca}

\end{document}